\documentstyle[aps,12pt,epsfig]{revtex}
\tightenlines

\def\<{\langle}
\def\>{\rangle}

\newcommand{\be}{\begin{equation}}
\newcommand{\ee}{\end{equation}}
\newcommand{\bea}{\begin{eqnarray}}
\newcommand{\eea}{\end{eqnarray}}

\begin{document}


\title{Finding cliques by quantum adiabatic evolution}
\author{Andrew M. Childs,$^1$ Edward Farhi,$^1$ Jeffrey Goldstone,$^1$ and
Sam Gutmann$^2$}
\address{$^1$ Center for Theoretical Physics, Massachusetts Institute of
              Technology, Cambridge, MA 02139 \\
         $^2$ Department of Mathematics, Northeastern University, Boston, 
	      MA 02115}
\date{19 December 2000}

\maketitle


\begin{abstract}
Quantum adiabatic evolution provides a general technique for the solution of
combinatorial search problems on quantum computers.  We present the results
of a numerical study of a particular application of quantum adiabatic
evolution, the problem of finding the largest clique in a random graph.  An
$n$-vertex random graph has each edge included with probability $1 \over 2$,
and a clique is a completely connected subgraph.  There is no known
classical algorithm that finds the largest clique in a random graph with
high probability and runs in a time polynomial in $n$.  For the small graphs
we are able to investigate ($n \le 18$), the quantum algorithm appears to
require only a quadratic run time.

\smallskip
\noindent MIT-CTP \#3067
\end{abstract}

\section{Introduction}

Quantum computation has been shown to have advantages in solving some
problems that require searching through a large space, but the exact nature
of this advantage remains an important open question.  In this paper, we
explore quantum adiabatic evolution, a general technique for solving such
problems.  Specifically, we consider the application of quantum adiabatic
evolution to the problem of finding the largest clique in a random graph.

Quantum adiabatic evolution provides a natural framework for solving
combinatorial search problems on quantum computers~\cite{FGGS00,KN98}.  The
Hamiltonian which governs the evolution of the quantum system interpolates
smoothly between an initial Hamiltonian whose ground state is easy to
construct and a final Hamiltonian whose ground state encodes the desired
solution.  The evolution of the quantum state proceeds in continuous time
according to the Schr\"odinger equation, starting in the ground state of the
initial Hamiltonian.  If the Hamiltonian varies slowly enough, the evolution
will closely track the instantaneous ground state and end in a state close
to the desired, final ground state.  Any problem which can be recast as the
minimization of an energy function (which can then be converted into a
quantum Hamiltonian) can potentially be solved in this way.  The key
question is how much time is required for the evolution to produce a final
state that gives a reasonable probability of finding the solution.

Previous work along these lines has focused on satisfiability problems, in
which the goal is to find an assignment of Boolean variables that makes a
certain logical expression over those variables true.  Early work
demonstrated that certain easy problems could in fact be solved efficiently
by adiabatic evolution~\cite{FGGS00}.  More general problems have been
treated numerically, and studies of a set of Exact Cover instances designed
to be hard have shown polynomial behavior out to instances containing as
many as twenty bits~\cite{FGG00,FGGLLP00}.  However, for satisfiability
problems like Exact Cover or 3-SAT, there are many ways to generate random
instances, and in general the observed performance of an algorithm depends
on the exact scheme chosen.

In this paper, we try to extend our understanding of the quantum adiabatic
evolution technique by studying its application to the problem of finding
the largest clique in a random graph.  There is a natural way to generate
random graphs, and for this distribution it is generally believed that no
polynomial-time classical algorithm will succeed in finding the largest
clique with high probability.  Thus an efficient quantum algorithm for this
problem would be an important step towards revealing the true power of
quantum computers.  Unfortunately, asymptotic analysis of quantum adiabatic
evolution algorithms appears to be difficult.

Here, we present a numerical study of our quantum adiabatic evolution
algorithm for finding cliques in graphs.  We first review adiabatic
evolution in general and discuss the properties of cliques in random graphs.
After showing how adiabatic evolution may be used to find cliques in any
graph, we present data showing that the median time required by the
algorithm to find the largest clique in a random graph apparently grows
quadratically for graphs of up to eighteen vertices.  We then focus on
graphs containing fifteen vertices and show that the algorithm behaves well
for the 8000 random graphs we generate.  It is possible that these results
on small graphs capture the asymptotic behavior of the algorithm, giving
some further evidence that quantum computation by adiabatic evolution may be
a good technique for solving hard combinatorial search problems.

\section{Quantum computation by adiabatic evolution}

Aside from measurements, a quantum system with the time-dependent
Hamiltonian $H(t)$ evolves according to the Schr\"odinger equation,
\be
  i {\partial \over \partial t} |\psi(t)\> = H(t) |\psi(t)\>
\label{eq:schrodinger}
\ee
(we set $\hbar=1$ throughout).  If $H(t)$ varies sufficiently slowly, and if
its instantaneous energy levels do not cross as a function of time, then the
quantum adiabatic theorem says that the evolution will track the
instantaneous eigenstates~\cite{Mes76}.  More specifically, suppose that we
wish to evolve from $t=0$ to $t=T$, the {\em run time}, and that we have a
one-parameter family of Hamiltonians $\tilde H(s)$ that varies smoothly for
$0 \le s \le 1$.  We set $H(t) = \tilde H(t/T)$ so that the run time $T$
governs how slowly $H$ varies.  Let
\be
  \tilde H(s) |j,s\> = E_j(s) |j,s\>
\ee
denote the instantaneous eigenstates of $\tilde H(s)$ with energy
eigenvalues $E_j(s)$ arranged in nondecreasing order.  Assume that $E_0(s)
\ne E_1(s)$ for all $0 \le s \le 1$, so that there is always a positive
energy gap between the ground and first excited states.  Time evolution
according to (\ref{eq:schrodinger}), starting in the initial ground state
$|\psi(0)\> = |0,s=0\>$, produces a final state $|\psi(T)\>$.  The adiabatic
theorem says that in the limit $T \to \infty$, $|\psi(T)\>$ is the final
ground state $|0,s=1\>$ (up to a phase).

Now imagine that the solution to an interesting computational problem can be
characterizing as minimizing a particular energy function.  This means we
can construct a Hamiltonian $H_P$ (the {\em problem Hamiltonian}) which is
diagonal in the computational basis and whose ground state encodes the
solution to the problem.  The quantum adiabatic theorem yields an idea for a
way to construct this ground state.  Suppose we have another Hamiltonian
$H_B$ (the {\em beginning Hamiltonian}) whose ground state --- perhaps a
uniform superposition over all possible solutions to the problem --- is easy
to construct.  If we choose the interpolating Hamiltonian
\be
  \tilde H(s) = (1-s) H_B + s H_P
\,,
\ee
so that
\be
  H(t) = \left(1-{t \over T}\right) H_B + {t \over T} H_P
\,,
\label{eq:interpolate}
\ee
then evolution from $t=0$ to $t=T$ starting in the ground state of $H_B$
will, in the adiabatic limit, produce the ground state of $H_P$, thus giving
the solution to the problem.

Of course, computation which takes an infinite amount of time is of little
practical value.  In practice, we would like to find a reasonably small
value of $T$ such that the final state gives us a reasonable chance of
finding the solution to the problem.  This time can be characterized in
terms of the spectrum of $\tilde H(s)$.  Let
\be
  g = \min_{0 \le s \le 1} \left( E_1(s)-E_0(s) \right)
\ee
denote the minimum gap over all values of $s$ between the ground state and
the first excited state, and let
\be
  {\cal E} = \max_{0 \le s \le 1} 
             |\<1,s|{\partial \tilde H \over \partial s}|0,s\>|
\ee
denote the most rapidly changing matrix element between the ground and first
excited state.  Then choosing
\be
  T \gg {{\cal E} \over g^2}
\ee
suffices to produce a final state arbitrarily close to the desired ground
state.  In typical problems of interest, $\cal E$ will scale polynomially
with the problem size, so the efficiency of the algorithm hinges on whether
$g$ is exponentially small or not.  Unfortunately, the size of this gap is
generally difficult to estimate analytically.

\section{Large cliques in random graphs}

Here, we review some simple graph-theoretic definitions.  For our purposes,
a {\em graph} $G$ is an $n \times n$ binary matrix that describes the
connectivity of a set of $n$ vertices labeled by the integers $1$ through
$n$.  The matrix element $G_{ij}$ is $1$ if vertices $i$ and $j$ are
connected by an edge and $0$ if they are not.  A {\em random graph} is a
graph in which each pair of vertices is connected, independently, with
probability $1 \over 2$.  A {\em clique} is a subgraph in which every pair
of vertices is connected by an edge.  In other words, $S \subseteq \{1,
\ldots, n\}$ is a clique in $G$ iff $G_{ij}=1$ for all $i,j \in S$, $i \ne
j$.

Many interesting properties of random graphs have been discovered since
their introduction by Erd\"os and R\'eyni~\cite{ER60}.  For a survey of such
results, see~\cite{Bol85}, and for a review of algorithms related to random
graphs, see~\cite{FM97}.  In particular, we are interested in algorithms for
finding large cliques in random graphs.  Roughly speaking, the largest
clique in a random graph with $n$ vertices has about $2 \log n$ vertices
(all logs are base 2).  In fact, given $n$, there is an integer $d(n)
\approx 2 \log n$ such that the largest clique has size $d(n)$ or $d(n)+1$
with probability tending to 1 as $n \to \infty$~\cite{Mat72}.

No polynomial time algorithm is known that will find, with high probability,
cliques of size $(1+\epsilon) \log n$ for any $\epsilon > 0$.  A simple
greedy heuristic will only produce cliques of size $1 \cdot \log n$ in
polynomial time.  Jerrum has analyzed in detail a more sophisticated
technique based on the Metropolis method, and he shows that it can require
super-polynomial time to find cliques larger than $\log n$~\cite{Jer92}.
Indeed, it has been conjectured that no efficient algorithm will find large
cliques, and this conjecture forms the basis of a proposed cryptographic
protocol~\cite{JP98}.  Our goal, then, is to investigate the possibility of
an efficient {\em quantum} algorithm which will find the largest clique in a
random graph.

\section{Algorithm}

We now present an algorithm based on quantum adiabatic evolution for finding
cliques in graphs.  This algorithm finds cliques of a particular size $k$.
Since random graphs asymptotically have a maximal clique with one of two
known sizes, it suffices to have a good algorithm for finding cliques of a
particular size.

The basis states in our Hilbert space will represent subsets of the set of
vertices $\{1,\ldots,n\}$.  Let the computational basis state $|z\> = |z_1
\ldots z_n\>$, where each $z_i=0$ or $1$, represent the subset which
includes vertex $i$ iff $z_i=1$.  Since we are only interested in subsets of
size $k$, we may restrict ourselves to the $n \choose k$-dimensional
subspace spanned by states $|z\>$ for which $h(z)=k$, where $h(z)$ denotes
the Hamming weight of $z$ (the number of ones that appear in its binary
representation $z_1 \ldots z_n$).

Our beginning Hamiltonian is
\be
  H_B = -\sum_{i>j}^n S^{ij}
\label{eq:hb}
\,,
\ee
where
\be
  S^{ij} = \left( \matrix{ 0 & 0 & 0 & 0 \cr
                           0 & 0 & 1 & 0 \cr
                           0 & 1 & 0 & 0 \cr
                           0 & 0 & 0 & 0} \right)^{ij}
\ee
acts on qubits $i$ and $j$ in the basis $\{|00\>,|01\>,|10\>,|11\>\}$.  Note
that $S^{ij}$ generates a swap between the $i$th and $j$th qubits.  In the
subspace of states of Hamming weight $k$ this Hamiltonian has the ground
state
\be
  |\psi_0\> = {n \choose k}^{-1/2} \sum_{h(z)=k} |z\>
\,,
\label{eq:initstate}
\ee
a uniform superposition of all states of Hamming weight $k$.  This is the
initial state for the algorithm.  It can be prepared efficiently from the
$|0\>$ state, as we describe in detail in the following section.

The problem Hamiltonian is diagonal in the computational basis:
\bea
  H_P |z\> = \sum_{i>j} (1-G_{ij}) z_i z_j |z\>
\,.
\label{eq:hp}
\eea
In other words, every pair of vertices that are in the state $|z\>$ but are
not connected in the graph raises the energy by one unit.  Thus the ground
state of this Hamiltonian (in the subspace of states of Hamming weight $k$)
will be a state with all $k$ vertices connected in the graph, assuming such
a state exists.

To summarize the algorithm, we prepare the system in the state given by
(\ref{eq:initstate}) and evolve according to the Hamiltonian
(\ref{eq:interpolate}), where $H_B$ is given by (\ref{eq:hb}) and $H_P$ is
given by (\ref{eq:hp}).  If there is a unique clique of size $k$, adiabatic
evolution will yield the corresponding state, which can easily be checked to
verify that it is indeed a clique.  If there are multiple cliques of size
$k$, we will find some superposition of the corresponding states, so that
measurement will give each of the various cliques with some probability.
Finally, if there is no clique of size $k$, we will instead find some subset
of $k$ vertices that maximizes the number of edges.

Our adiabatic algorithm is naturally defined in continuous time.  However,
since the Hamiltonian is a sum of polynomially many two-qubit operations,
the evolution operator can be approximated by a product of two-qubit unitary
operators with polynomial overhead~\cite{FGGS00}.

\section{Preparing the initial state}

There are many ways to efficiently prepare the initial state
(\ref{eq:initstate}).  Directly computing the state is possible, but we do
not know of any particularly straightforward method.  However, it can be
easily prepared using projective measurements.  Starting in the $n$-qubit
state $|0\>$, we apply the biased Hadamard transform
\be
  \left( \matrix{ \sqrt{1-{k \over n}} & \sqrt{k \over n} \cr
                  \sqrt{k \over n}     & -\sqrt{1-{k \over n}} } 
  \right)^{\otimes n}
\label{eq:biashad}
\ee
and measure the Hamming weight.  Note that the Hamming weight can be
efficiently measured by performing addition of each of the $n$ qubits into
an ancilla of size $\log n$ initialized to the $|0\>$ state~\cite{CM00}.
Measuring the ancilla in the computational basis then gives a measurement of
the Hamming weight.  Since both the initial state and the measurement
outcome are invariant under interchange of any two qubits, this measurement
will produce a uniform superposition of states with Hamming weight given by
the measurement outcome.

The state produced by (\ref{eq:biashad}) has a binomial distribution of
Hamming weights with mean $k$, so the probability of the measurement
yielding this mean is
\be
  p(n,k) = {n \choose k} \left({k \over n}\right)^k 
           \left(1-{k \over n}\right)^{n-k}
\,.
\ee
For fixed $n$, this function has a minimum at $k={n \over 2}$, at which
point $p(n,n/2) \approx {\sqrt {2 \over n \pi}}$.  Thus $p(n,k) \gtrsim
{\sqrt {2 \over n \pi}}$ independent of $k$, and hence the we only need to
repeat the procedure polynomially many times to produce a state with Hamming
weight $k$.

It is also possible to produce a state arbitrarily close to
(\ref{eq:initstate}) by adiabatic evolution.  No measurements are required,
and it is easy to understand how the method works.  We take the beginning
Hamiltonian
\be
  H_B^0 = {1 \over 2} \sum_i (1-\sigma_x^i)
\,,
\ee
where
\be
  \sigma_x^i = \left( \matrix{ 0 & 1 \cr
                               1 & 0 } \right)^i
\ee
is the Pauli $x$ operator on the $i$th qubit.  This Hamiltonian has the
ground state
\be
  |\psi^0_0\> = 2^{-n/2} \sum_z |z\>
\,,
\ee
a uniform superposition of computational basis states, which can be easily
prepared by Hadamard transformation of the $|0\>$ state.  We take the
problem Hamiltonian defined by
\be
  H_P^0 |z\> = \left( \sum_i z_i-k \right)^2 |z\>
\,.
\ee
The ground states of $H_P^0$ are the states of Hamming weight $k$.  By
symmetry, the final state achieved by adiabatic evolution will be close to
(\ref{eq:initstate}).

The Hamiltonian for this problem is particularly simple because it depends
only on the total spin in the $x$ and $z$ directions.  If we let $S_a$
denote the total spin in the $a$ direction (where $a \in \{x,y,z\}$), then
we have
\bea
  H_B^0 &=& {n \over 2} - S_x \\
  H_P^0 &=& \left({n \over 2} - S_z - k\right)^2
\,.
\eea
The Hamiltonian commutes with $\vec S^2 = \sum_a S_a^2$, and the initial
state is symmetric, so we may work in the $(n+1)$-dimensional subspace of
symmetric states, those with $\vec S^2={n \over 2}({n \over 2}+1)$, and
choose as basis states the eigenstates $|m\>$ of $S_z$ satisfying
\be
  S_z |m\> = m |m\>, \quad m=-{n \over 2}, -{n \over 2}+1, 
                     \ldots, {n \over 2}
\,.
\ee
Using the matrix elements
\be
  \<m'|S_x|m\> \nonumber\\ = {1 \over 2}
    \left(\sqrt{{n \over 2}\left({n \over 2}+1\right)-m(m+1)}
           \,\delta_{m,m'-1}        
	 +\sqrt{{n \over 2}\left({n \over 2}+1\right)-m'(m'+1)}
           \,\delta_{m',m-1} \right)
\,,
\ee
we may easily show numerically that for large $n$, the minimum gap occurs
near $s=1$, so that the minimum gap is one independent of $n$.  Thus a
polynomially large $T$ suffices to produce a state arbitrarily close to
(\ref{eq:initstate}).

\section{Results}

To study the behavior of our algorithm for finding a clique of size $k$ in a
randomly generated graph, we numerically integrate the Schr\"odinger
equation (\ref{eq:schrodinger}) starting from the initial state
(\ref{eq:initstate}).  We use a fifth-order Runge-Kutta integrator with
adaptive step size.  Although the quantum system representing an $n$-vertex
graph can be thought of as living in a $2^n$-dimensional Hilbert space, we
are interested only in the subspace of states of Hamming weight $k$, which
reduces the problem to an $n \choose k$-dimensional subspace.

For our simulation to run in a reasonable amount of classical computer time,
we choose some fixed probability of success as our goal, where ``success''
means that a measurement of the final state in the computational basis
yields a clique of size $k$.  We choose a success probability of $1 \over
8$, which is significantly higher than $2^{-n}$ for the cases of interest
but gives run times that are not too long.  For each random graph generated,
we determine how long the algorithm must run so that the probability of
finding a clique of the desired size upon measurement of the final state is
$1 \over 8$.  Note that any fixed probability (independent of $n$) can be
made exponentially close to unity by polynomially many repetitions.  

Initially, we consider only the set of random graphs with unique maximal
cliques.  It seems intuitive that finding the maximal clique should be
hardest in this case, a conjecture that is borne out by later results.  We
concentrate on this more specific set to get tighter statistics and thus a
better picture of the behavior of the algorithm at the numbers of bits we
are able to investigate.

After generating a random graph with $n$ vertices, we classically determine
the size $k$ of the largest clique in the graph.  This is easy because $n
\le 18$.  Whatever $k$ is, we then attempt to find, by (simulated) quantum
adiabatic evolution, a clique of size $k$.  In the interest of generating
smooth statistics and discovering the true asymptotic performance of the
algorithm, we simply average over the different values of $k$ that appear,
weighted by their frequency of occurrence.

\begin{figure}
\begin{center}
\psfig{file=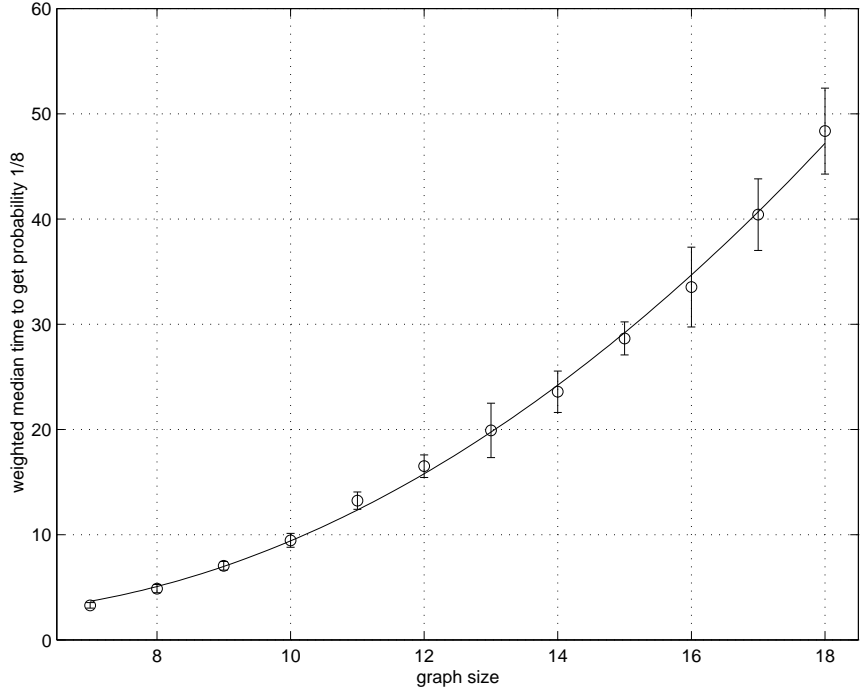,width=4.5in}
\end{center}
\caption{Median times to get success probability $1 \over 8$ for random
graphs with unique maximal cliques.  Error bars show the 95\% confidence
level on the median.}
\label{fig:median}
\end{figure}

For each $n$, $7 \le n \le 18$, we generated 100 random graphs of size $n$
with unique maximal cliques.  Fig.~\ref{fig:median} shows the median time to
achieve a success probability ${1 \over 8} \pm {1 \over 400}$ of finding the
clique of maximal size.  The solid line is a fit to a quadratic,
$T(n)=0.255\, n^2 - 2.43\, n + 8.15$.  The good fit to a quadratic suggests
that the median run time to get probability $1 \over 8$ may be a polynomial
function of the graph size.

\begin{figure}
\begin{center}
\psfig{file=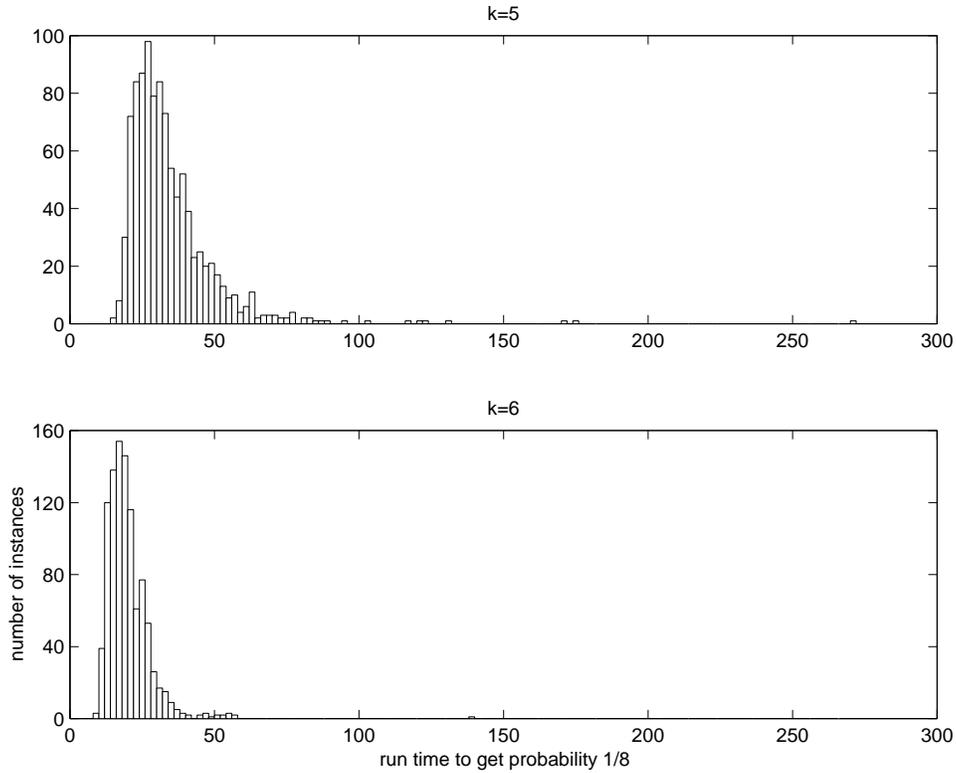,width=5in}
\end{center}
\caption{Distributions of the run times to get probability $1 \over 8$ for
2000 random graphs of size $n=15$ with unique maximal cliques of sizes $k=5$
or $k=6$.  Note that the most outlying points are near $T=270$ for $k=5$ and
$T=140$ for $k=6$.}
\label{fig:timehist}
\end{figure}

Although Fig.~\ref{fig:median} captures how the algorithm's performance
scales with $n$, and the error bars suggest that the distribution of run
times is not too broad, we would like to understand more detailed features
of the distribution.  We choose to focus on graphs of size $n=15$, as this
size is at the edge of our capability to simulate a large number of
instances in a reasonable amount of time.  At $n=15$, the random graphs with
unique maximal cliques are most likely to have a maximal clique size of
either $k=5$ or $k=6$, so we consider only these two values.  We begin by
accurately determining the median run time to get probability $1 \over 8$ by
generating 1000 instances with unique maximal cliques at each of $k=5$ and
$k=6$.  We find median run times $T_5 = 30.87$ and $T_6 = 18.56$.  The full
distributions of run times for these instances are shown in
Fig.~\ref{fig:timehist}.  Note that unique maximal cliques of size $k=6$ are
found faster than those of size $k=5$.

To specify a general algorithm for finding cliques in graphs of arbitrary
size, we must provide a procedure for choosing the run time $T$ at any value
of $n$.  We might imagine a procedure wherein we start at some
$n$-independent $T$ and repeatedly increase $T$ in some way if the algorithm
fails to find a clique.  However, if the median time to achieve some fixed
probability is truly asymptotically quadratic, and if the distribution about
this median is not too broad, a reasonable procedure is to simply choose a
run time by extrapolating the fit shown in Fig.~\ref{fig:median}.

\begin{figure}
\begin{center}
\psfig{file=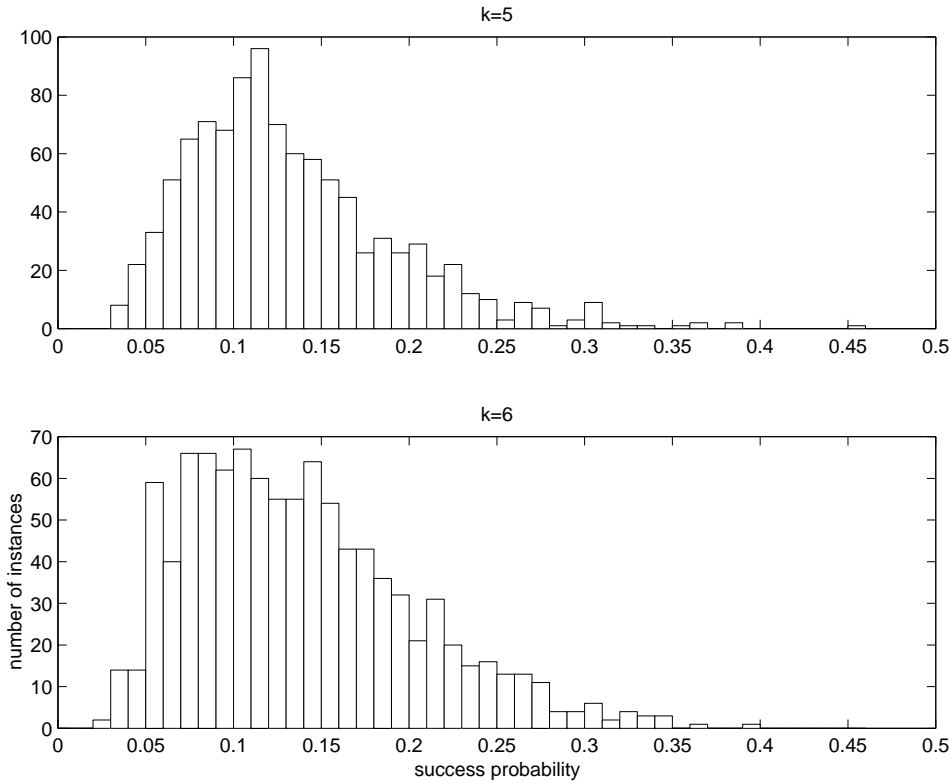,width=5in}
\end{center}
\caption{Distributions of success probabilities at the median run time for
2000 random graphs of size $n=15$ with unique maximal cliques of sizes $k=5$
or $k=6$.}
\label{fig:uprobhist}
\end{figure}

In view of the latter approach, having determined the median run times for
$n=15$, we would like to know the distribution of success probabilities at
these run times.  These distributions are shown in Fig.~\ref{fig:uprobhist}
for $k=5$ and $k=6$, again with 1000 instances at each $k$ and constraining
the graphs to have unique maximal cliques.  Unsurprisingly, we find median
success probabilities near $1 \over 8$: $p_5=0.120$ and $p_6=0.128$.  More
importantly, the distribution of success probabilities appears to be cut off
fairly sharply on the low probability side.  Indeed, we find minimum
probabilities $p_5^{\rm min}=0.031$ and $p_6^{\rm min}=0.023$.

\begin{figure}
\begin{center}
\psfig{file=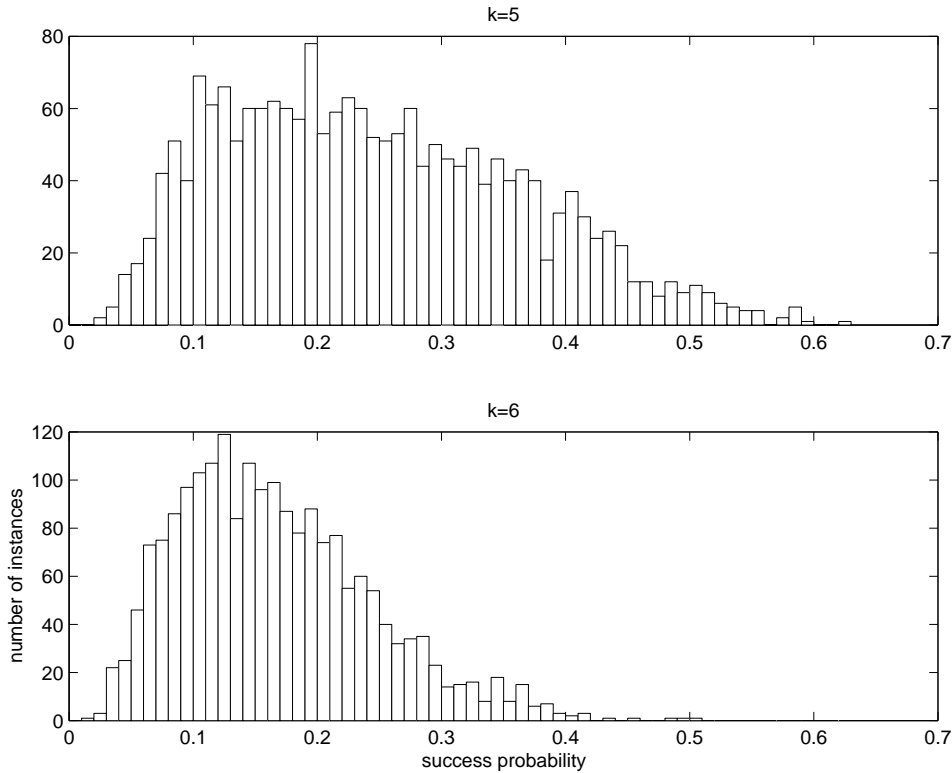,width=5in}
\end{center}
\caption{Distribution of success probabilities at the median run time for
4000 random graphs of size $n=15$ with (not necessarily unique) maximal
cliques of sizes $k=5$ or $k=6$.}
\label{fig:nonuprobhist}
\end{figure}

Ultimately, we are interested in the distribution of success probabilities
over all random graphs, with or without a unique maximal clique.
Fig.~\ref{fig:nonuprobhist} shows the distribution of success probabilities
without the constraint that the maximal clique is unique, based on 2000
instances at each of $k=5$ and $k=6$.  These distributions are shifted
towards higher probabilities than in the unique case, with median
probabilities $p_5'=0.232$ and $p_6'=0.155$, both higher than $1 \over 8$.
The distributions still fall off sharply on the low probability side.  Thus,
it is clear that finding non-unique maximal cliques is easier for the
quantum algorithm than finding unique ones, and we are justified in having
determined the run time using graphs for which the maximal clique is unique.

\section{Conclusion}

We have presented data showing that quantum computation by adiabatic
evolution is a reasonable candidate for a fast algorithm to find the largest
clique in a random graph.  Together with previous studies of the performance
of similar methods for solving satisfiability problems, these results
suggest that quantum computation by adiabatic evolution may be a useful,
general way to attack difficult combinatorial search problems.

\section*{Acknowledgements}

We thank David Beckman and Leslie Valient for several helpful discussions.
We also thank the MIT Laboratory for Nuclear Science Computer Facility for
use of the computer Abacus.  This work was supported in part by the
Department of Energy under cooperative agreement DE-FC02-94ER40818.  AMC is
supported by the Fannie and John Hertz Foundation.



\begin{thebibliography}{10}
\bibitem{FGGS00}
  E. Farhi, J. Goldstone, S. Gutmann, and M. Sipser, {\em Quantum
  computation by adiabatic evolution}, quant-ph/0001106.
\bibitem{KN98}
  For a related technique, see T. Kadowaki and H. Nishimori, {\em Quantum
  annealing and the transverse Ising model}, cond-mat/9804280; Phys. Rev. E
  {\bf 58}, 5355 (1998).
\bibitem{FGG00}
  E. Farhi, J. Goldstone, and S. Gutmann, {\em A numerical study of the
  performance of a quantum adiabatic evolution algorithm for
  satisfiability}, quant-ph/0007071.
\bibitem{FGGLLP00}
  E. Farhi, J. Goldstone, S. Gutmann, J. Lapan, A. Lundgren, and D. Preda,
  {\em A quantum adiabatic evolution algorithm applied to an NP-complete
  problem}, to appear.
\bibitem{Mes76}
  A. Messiah, {\em Quantum Mechanics}, Vol. II (Wiley, New York, 1976).
\bibitem{ER60}
  P. Erd\"os and A. R\'eyni, {\em On the evolution of random graphs}, Publ.
  Math. Inst. Hung. Acad. Sci. {\bf 8}, 455 (1964).
\bibitem{Bol85}
  B. Bollob\'as, {\em Random Graphs} (Academic, New York, 1985).
\bibitem{FM97}
  A. Frieze and C. McDiarmid, {\em Algorithmic theory of random graphs},
  Random Struct. Alg. {\bf 10}, 5 (1997).
\bibitem{Mat72}
  D. W. Matula, {\em The employee party problem}, Notices Amer. Math. Soc.
  {\bf 19}, A-382 (1972).
\bibitem{Jer92}
  M. Jerrum, {\em Large cliques elude the Metropolis process}, Random Struct.
  Alg. {\bf 3}, 347 (1992).
\bibitem{JP98}
  A. Juels and M. Peinado, {\em Hiding cliques for cryptographic security},
  Proc. 9th Annual ACM-SIAM SODA, 678 (1998).
\bibitem{CM00}
  I. L. Chuang and D. S. Modha, {\em Reversible arithmetic coding for
  quantum data compression}, IEEE Trans. Inf. Theory {\bf 46}, 1104 (2000).
\end{thebibliography}
\end{document}